\documentclass[aps,floatfix,nofootinbib,onecolumn,superscriptaddress]{revtex4}
\pdfoutput=1

\usepackage{amsmath}
\usepackage{amssymb}
\usepackage{amsthm}
\usepackage{dcolumn}
\usepackage{epsfig}
\usepackage{graphics}
\usepackage{graphicx}
\usepackage{slashed,epsfig}
\usepackage{longtable}
\usepackage{color}

\definecolor{darkgreen}{rgb}{0,0.5,0}
\definecolor{purple}{rgb}{0.5,0,0.5}
\definecolor{nblue}{rgb}{0.0,0.0,0.50}
\definecolor{scarlet}{rgb}{1.0,0.2,0}
\usepackage[colorlinks=true, pdfstartview=FitV, linkcolor=purple, citecolor= purple, urlcolor=blue]{hyperref}

\newcommand{\beq} {\begin{equation}}
\newcommand{\eeq} {\end{equation}}
\newcommand{\beqa} {\begin{eqnarray}}
\newcommand{\eeqa} {\end{eqnarray}}

\newcommand{\mbf}[1] {{\mathbf{#1}}}

\begin{document}

{\par\raggedleft \texttt{SLAC-PUB-14865}\par}
\bigskip{}

\title{Novel Perspectives for Hadron Physics}

\author{Stanley~J.~Brodsky} \affiliation{SLAC National Accelerator Laboratory\\
Stanford University, Stanford, California 94309, USA}

\begin{abstract}

I discuss several novel and unexpected aspects of quantum chromodynamics.  These include: (a) the nonperturbative origin of intrinsic strange, charm and bottom quarks in the nucleon at large $x$;  the  breakdown of pQCD factorization theorems due to the lensing effects of initial- and final-state interactions; (b) important corrections to pQCD scaling for inclusive reactions due to processes in which hadrons are created at high transverse momentum directly in the hard processes and their relation to the baryon anomaly in high-centrality heavy-ion collisions; and (c) the nonuniversality of quark distributions in nuclei.  I also discuss some novel theoretical perspectives in QCD: (a) light-front holography -- a relativistic color-confining first approximation to QCD based on the AdS/CFT correspondence principle; (b) the principle of maximum conformality -- a method which determines the renormalization scale at finite order in perturbation theory yielding scheme independent results; (c) the replacement of quark and gluon vacuum condensates by ``in-hadron condensates"  and how this helps to resolves the conflict between QCD vacuum and the cosmological constant.

\end{abstract}


\maketitle

\date{\today}

\section{Introduction}
\label{intro}

One of the most remarkable achievements in the history of science was the development~\cite{Fritzsch:1973pi} of quantum chromodynamics, the renormalizable gauge theory of color-triplet quark and color-octet gluon fields.   QCD is believed to be the  fundamental theory  of hadron and nuclear phenomena in the same sense that  quantum electrodynamics provides the fundamental theory underlying  all of atomic physics and chemistry.  In fact,  quantum electrodynamics  can be regarded as the zero-color limit of quantum chromodynamics~\cite{Brodsky:1997jk}

QCD predictions based on the nearly scale-invariant interactions of  quarks and gluons at short distances and asymptotic freedom have been validated by many measurements,  such as deep inelastic lepton scattering,  electron-positron annihilation into hadrons, and quark and gluon jet production in high energy hadronic collisions.   However,  phenomena in the nonperturbative color-confining strong-coupling domain can be extraordinarily complex and can have unexpected features.

In this talk I will review a number of unexpected features of quantum chromodynamics,  especially the novel effects arising from the heavy-quark quantum fluctuations of hadron wavefunctions. I will also  discuss corrections to pQCD leading-twist scaling for inclusive reactions due to processes in which hadrons are created at high transverse momentum directly in the hard process; the baryon anomaly in high centrality heavy ion collisions; the  breakdown of  factorization theorems due to the lensing effects of initial- and final-state interactions; and the non-universality of quark distributions in nuclei.  I will also discuss some novel theoretical perspectives in QCD: (a) light-front holography -- a first approximation to QCD based on the AdS/CFT correspondence principle; (b) the principle of maximum conformality -- which determines the renormalization scale order-by-order in perturbation theory yielding scheme independent results; (c) the replacement of quark and gluon vacuum condensates by ``in-hadron condensates"  and how this resolves the conflict between the physics of the QCD vacuum and the  cosmological constant.    QCD also predicts a rich array of novel hadronic and nuclear phenomena.  These include the production of a quark-gluon plasma in high energy, high density heavy ion collisions, ``color transparent"  interactions of hadrons in nuclear reactions,   and ``hidden-color"  degrees of freedom in nuclei.

\section{Intrinsic Heavy Quarks}

If one follows conventional wisdom, nonvalence ``sea " quarks in the proton structure functions only arise from gluon splitting $g \to Q \bar Q; $ i.e.,  the proton wavefunction at an initial soft scale is assumed to only contain valence quarks and gluons.  DGLAP evolution from the $g  \to Q \bar Q$ spitting process is then assumed to  generate  all of the sea quarks at virtuality $Q^2 > 4 m^2_Q$.  If this hypothesis were correct, then the $\bar u(x)$ and $\bar d(x)$ distributions would be identical.    Similarly, if sea quarks only arise from gluon splitting, one expects that the $s(x) $and $\bar s(x)$ distributions will be the same and fall-off faster in $x$ than the parent gluon distributions.  However, measurements of Drell-Yan processes, deep inelastic electron and neutrino scattering, and other experiments show that these simplified predictions are incorrect.

\begin{figure}
 \begin{center}
\includegraphics[width=16cm]{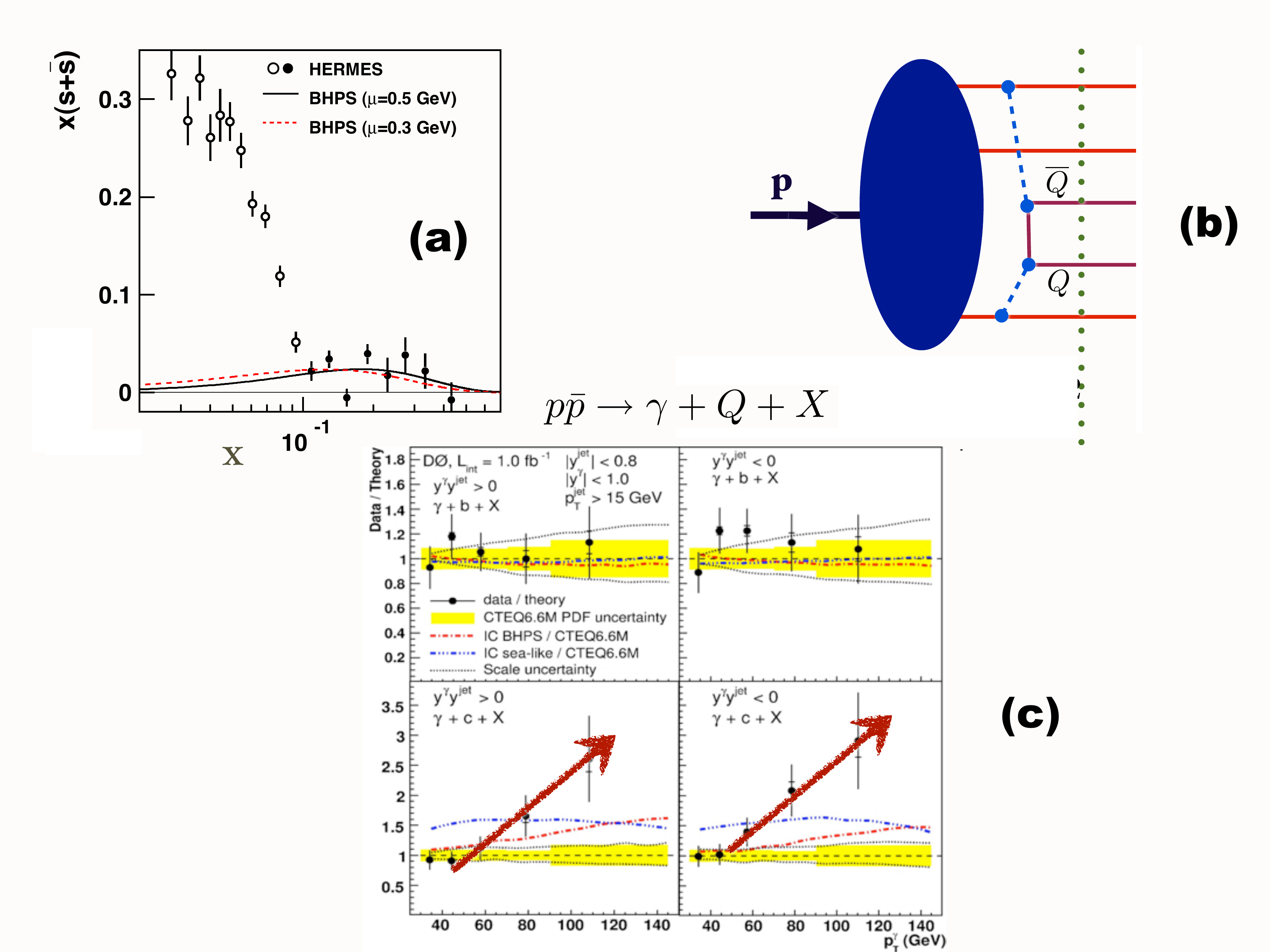}
\end{center}
\caption{(a) Intrinsic and extrinsic strangeness distribution.~\cite{Chang:2011du}
(b) Five-quark Fock state of the proton and the origin of the intrinsic sea.
(c) D0 measurement of $ \bar p p \to \gamma + b X$ and $ \bar p p \to \gamma + c X.$}
\label{Hermes}
\end{figure}

The five-quark Fock state of the proton's LFWF $ \vert uud Q \bar Q \rangle$ is the primary origin of the sea-quark distributions of the proton.  Experiments show that the sea quarks have remarkable nonperturbative features, such as $\bar u(x) \ne \bar d(x)$, and an intrinsic strangeness~\cite{Airapetian:2008qf} distribution $s(x)$ appearing at light-cone momentum fraction $x > 0.1$, as well as intrinsic charm and bottom distributions at large $x$.
In fact, recent measurements from  HERMES show that the strange quark in the proton has two distinct components: a fast-falling contribution consistent with gluon splitting to $s \bar s$ and an approximately flat component up to $ x < 0.5$. See fig. \ref{Hermes}(a).

The proton light-front wavefunction in QCD contains {\it ab initio } intrinsic heavy-quark Fock state components such as $\vert uud c \bar c\rangle$.~ \cite{Brodsky:1980pb,Brodsky:1984nx,Harris:1995jx,Franz:2000ee} Such distributions~\cite{Brodsky:1984nx,Franz:2000ee} favor configurations where the quarks have equal rapidity.  The intrinsic heavy quarks thus carry most of the proton's momentum since this minimizes the off-shellness of the state. These configurations  arise, for example, from $g g \to Q \bar Q \to gg $ insertions connected to the valence quarks in the proton self-energy; See  Fig. \ref{Hermes}(b). in fact, the intrinsic strangeness, charm and $\bar u(x)-\bar d(x)$ distributions fit a universal intrinsic quark model,~\cite{Brodsky:1980pb} as recently shown by Chang and Peng.~\cite{Chang:2011du}.
QCD also predicts that the heavy quark pair $Q \bar Q$ in the intrinsic five-quark Fock state  is primarily a color-octet,  and the ratio of intrinsic charm to intrinsic bottom scales
as $m_c^2/m_b^2 \simeq 1/10,$ as can easily be seen from the operator product expansion in non-Abelian QCD.~\cite{Brodsky:1984nx,Franz:2000ee}  Intrinsic charm and bottom thus can explain the origin of high open-charm and open-bottom hadron production at high momentum fractions, as well as the single and double $J/\psi$ hadroproduction cross sections observed at high $x_F$.

In the case of a hadronic high energy proton collision, the high-$x$ intrinsic charm quark in  the proton's $|uud c \bar c>$ Fock state can   coalesces with the co-moving $ud$ valence quarks in a projectile proton to produce a  $\Lambda_c(cud)$ baryon at the combined high  momentum fraction $x_F = x_u + x_d + x_c$.
Similarly,  the coalescence of comoving $b$ and $\bar u$ quarks from the $|uud \bar b b>$  intrinsic bottom Fock state explains the production of the $\Lambda_b(udb)$  which was first observed at the ISR collider at CERN by Cifarelli, Zichichi, and their collaborators~\cite{Bari:1991ty}.  Furthermore, one finds that the $\Lambda_{b}$  is produced in association with a positron from the decay of the associated high-$x_F$ $B^0(u \bar b)$ meson.

As emphasized by Lai, Tung, and Pumplin~\cite{Pumplin:2007wg}, the structure functions used to model charm
and bottom quarks in the proton at large $x_{bj}$ have been consistently underestimated, since they ignore intrinsic heavy quark fluctuations of
hadron wavefunctions.
Furthermore, the neglect of the intrinsic-heavy quark component in the proton structure function will lead to an incorrect assessment of the gluon distribution at large $x$ if it is assumed that sea quarks always arise from gluon splitting~\cite{Stavreva:2010mw}

The D0 collaboration~\cite{Abazov:2009de} at the Tevatron has recently measured the processes $\bar p p \to c + \gamma + X$ and $\bar p p \to b + \gamma + X$ at very high photon transverse momentum: $p_T^\gamma \sim 120~{\rm GeV/c}$.  As seen in  Fig. \ref{Hermes}(c),
the rate for $\bar p p \to b + \gamma  X$ for bottom quark jets agrees very well with NLO PQCD predictions; however the corresponding charm jet cross section deviates strongly from the standard PQCD prediction for  $p_T^\gamma > 60~ GeV/c$.  This  photon plus charm jet anomaly can be explained if one allows for an intrinsic contribution to the charm structure function in $g c \to c \gamma$  at $Q^2 \sim 10^4~{\rm GeV}^2$, but it requires a factor of two increase in strength compared to the CTEQ parameterization. This discrepancy could indicate that the reduction of the charm distribution due to DGLAP evolution has been overestimated.

The SELEX collaboration~\cite{Engelfried:2005kd} has reported the discovery of a set of  doubly-charmed spin 1/2 and spin 3/2 baryons with quantum numbers matching
$|ccu \rangle$ and   $|ccd \rangle $ bound states.  The NA3 experiment has also observed the hadroproduction of two $J/\psi$s at high $x_F$, also a signal for seven quark Fock states like $|uud c \bar c c\bar c>$  in the proton. However, the mass splittings of the $ccu$ and $ccd$ states measured by SELEX  are much larger than expected from known QCD isospin-splitting mechanisms.
One  speculative proposal~\cite{Brodsky:2011zs} is that these baryons have a linear configuration $c~q~c$ where the light quark $q$ is exchanged between the heavy quarks as in a linear molecule.  The linear  configuration enhances the Coulomb repulsion of the $c~u~c$ relative to $c~d~c$.  It is clearly important to have experimental confirmation of the SELEX results.

\begin{figure}
 \begin{center}
\includegraphics[width=8cm]{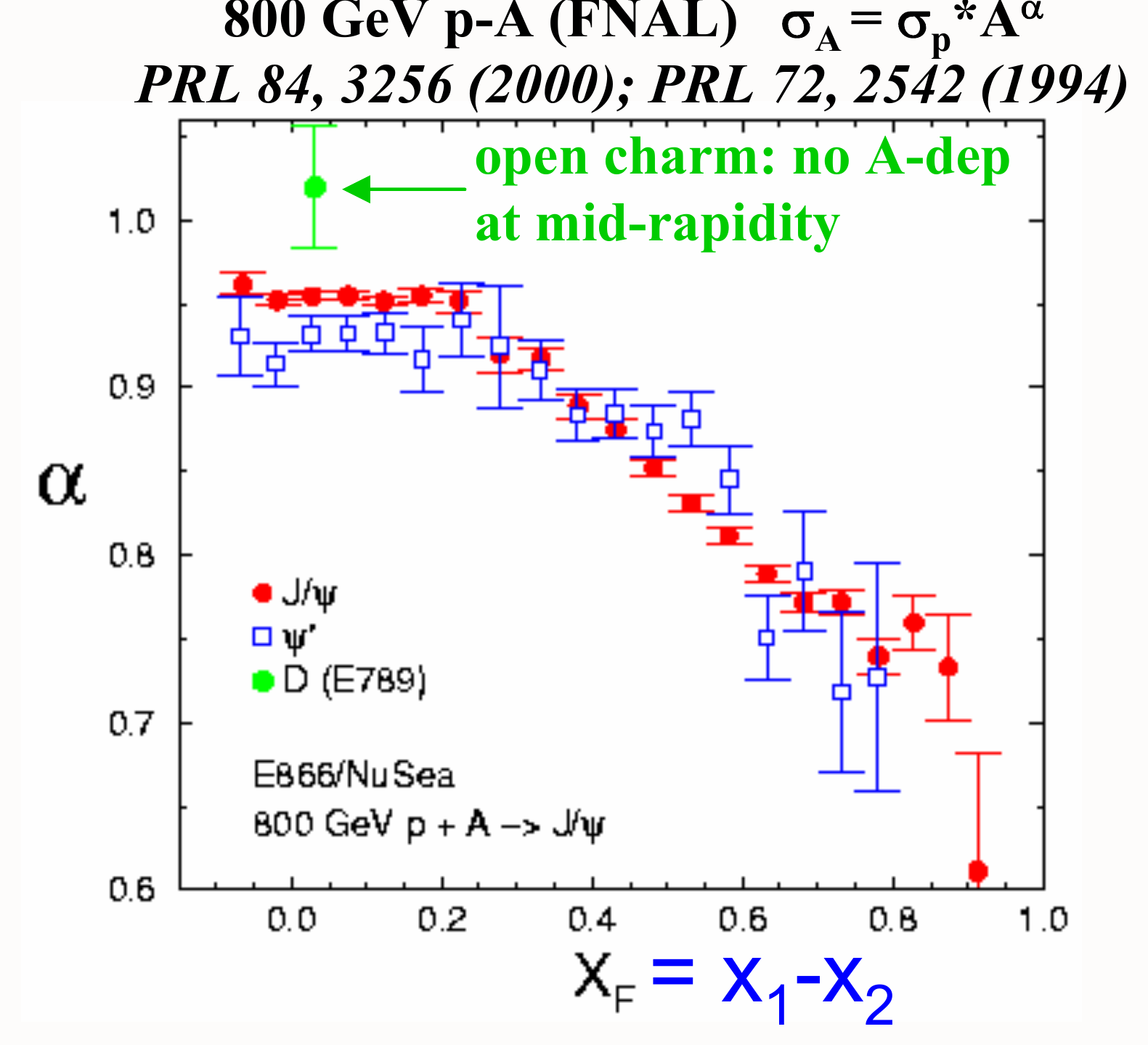}
\includegraphics[width=8cm]{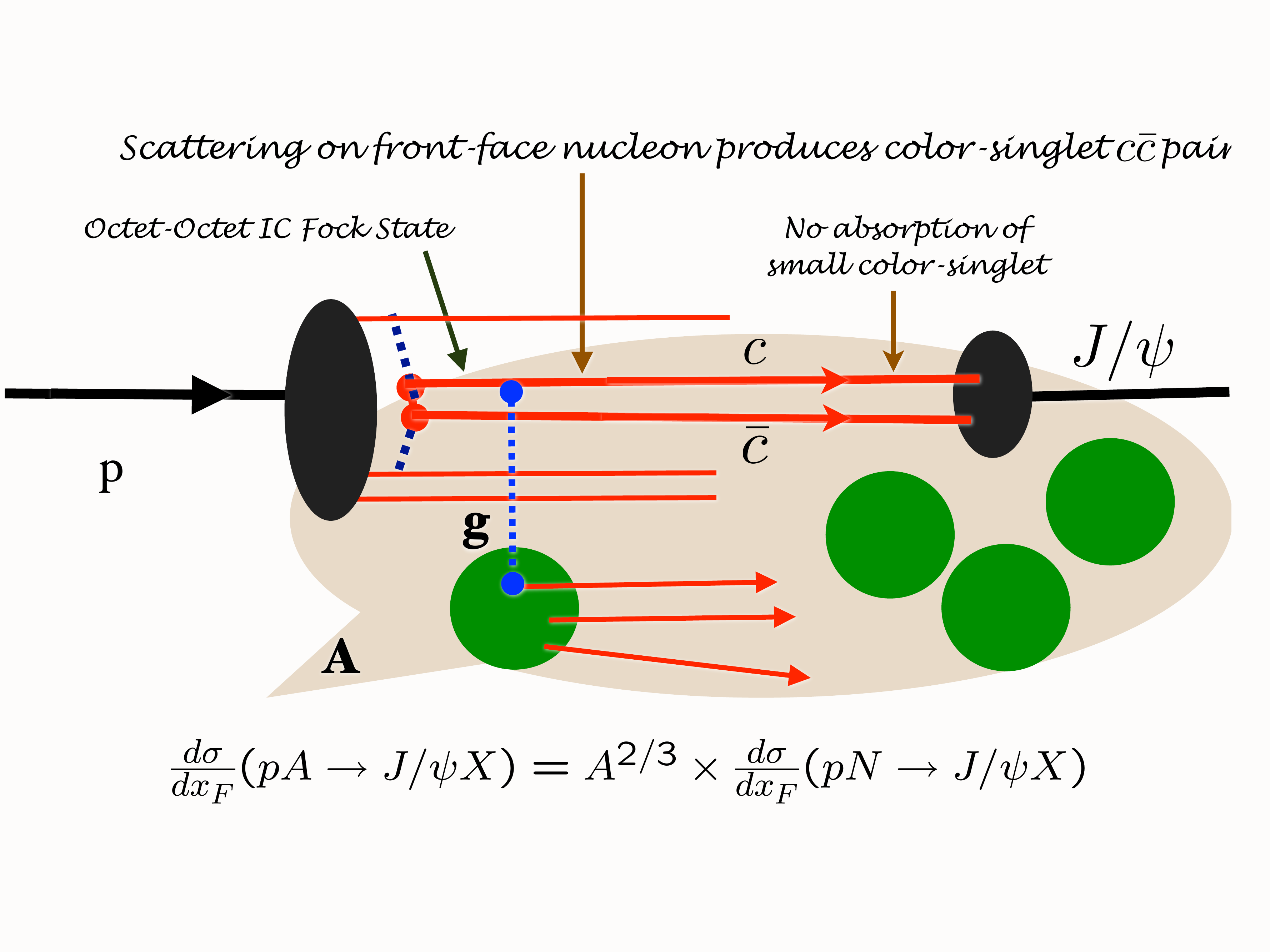}
\end{center}
\caption{(a) E866/NuSea data for the nuclear $A$ dependence of $J/\psi$ and $\psi^\prime$ hadroproduction. (b) Model for the $A$ dependence of $J/\psi$ hadroproduction based on color-octet intrinsic charm.}
\label{AdepJpsi}
\end{figure}

The cross section for $J/\psi$ production in a nuclear target is well measured. The ratio of the nuclear and proton target cross sections has the form $A^{\alpha(x_F)}$ where $x_F$ is Feynman fractional longitudinal momentum of the $J/\psi$. At small $x_F$, $\alpha(x_F)$  is slightly smaller than one but at $x_F \sim 1$ it decreases to $\alpha=2/3$. These results, as shown in  Fig.~\ref{AdepJpsi}(a), are surprising since (1)  the value $\alpha= 2/3$ would be characteristic of a strongly interacting hadron, not a small-size quarkonium state; and (2) the functional dependence   $A^{\alpha(x_F)}$ contradicts  pQCD factorization predictions.
This anomaly, in combination with the anomalously large and flat cross sections measured at high $x_F$, is consistent with a QCD mechanism based on color-octet intrinsic charm Fock states: because of its large color dipole moment, the intrinsic heavy quark Fock state of the proton: $|(uud)_{8_C} (c \bar c)_{8_C} \rangle$ interacts primarily with the $A^{2/3} $nucleons at the front surface.  See Fig.~\ref{AdepJpsi}(b).
The $c \bar c$ color octet thus scatters on a front-surface nucleon, changes to a color singlet, and then propagates through the nucleus as a $J/\psi$ at high $x_F$.   Alternatively, one can postulate strong energy losses of a color octet $c \bar c$ state as it propagates in the nucleus but it is hard to see how this can account for the observed nearly flat behavior of the $A^{2/3}$ component as observed by NA3.

Intrinsic heavy quarks also provide a novel mechanism for the inclusive and diffractive
Higgs production $pp \to p p H$, in which the Higgs boson carries a significant fraction of the projectile proton momentum.~\cite{Brodsky:2006wb,Brodsky:2007yz}  The production
mechanism is based on the subprocess $(Q \bar Q) g \to H $ where the Higgs acquires the momentum of the $Q \bar Q$ pair in the $\vert uud Q \bar Q \rangle$ intrinsic heavy quark Fock state of the colliding proton and thus has approximately
$80\%$ of the projectile proton's momentum.   The high-$x_F$ Higgs could be accessed at the LHC using far forward detectors or arranging the proton beams to collide at a significant crossing angle.  It is also possible to produce a light mass  Higgs at threshold using the $7$ TeV proton beam colliding with  a fixed nuclear target.

\section{The Unexpected Role of Direct Processes in High $p_T$ Hadron Reactions}

It is normally assumed that hadrons produced at high transverse momentum  in inclusive high energy hadronic collisions  such as $ p p \to H X$ only arise  from quark and gluon jet fragmentation.
A  fundamental test of leading-twist QCD predictions in high transverse momentum hadronic reactions is the measurement of the power-law
fall-off of the inclusive cross section~\cite{Sivers:1975dg}
${E d \sigma/d^3p}(A B \to C X) ={ F(\theta_{cm}, x_T)/ p_T^{n_{\rm eff}} } $ at fixed $x_T = 2 p_T/\sqrt s$
and fixed $\theta_{CM}$. In the case of the scale-invariant  parton model  $n_{\rm eff} = 4.$    However in QCD $n_{\rm eff} \sim 4 + \delta$ where  $\delta \simeq 1.5 $  is the typical correction to the conformal prediction arising
from the QCD running coupling and the DGLAP evolution of the input parton distribution and fragmentation
 functions.~\cite{Arleo:2009ch,Arleo:2010yg}

\begin{figure}
 \begin{center}
\includegraphics[width=8cm]{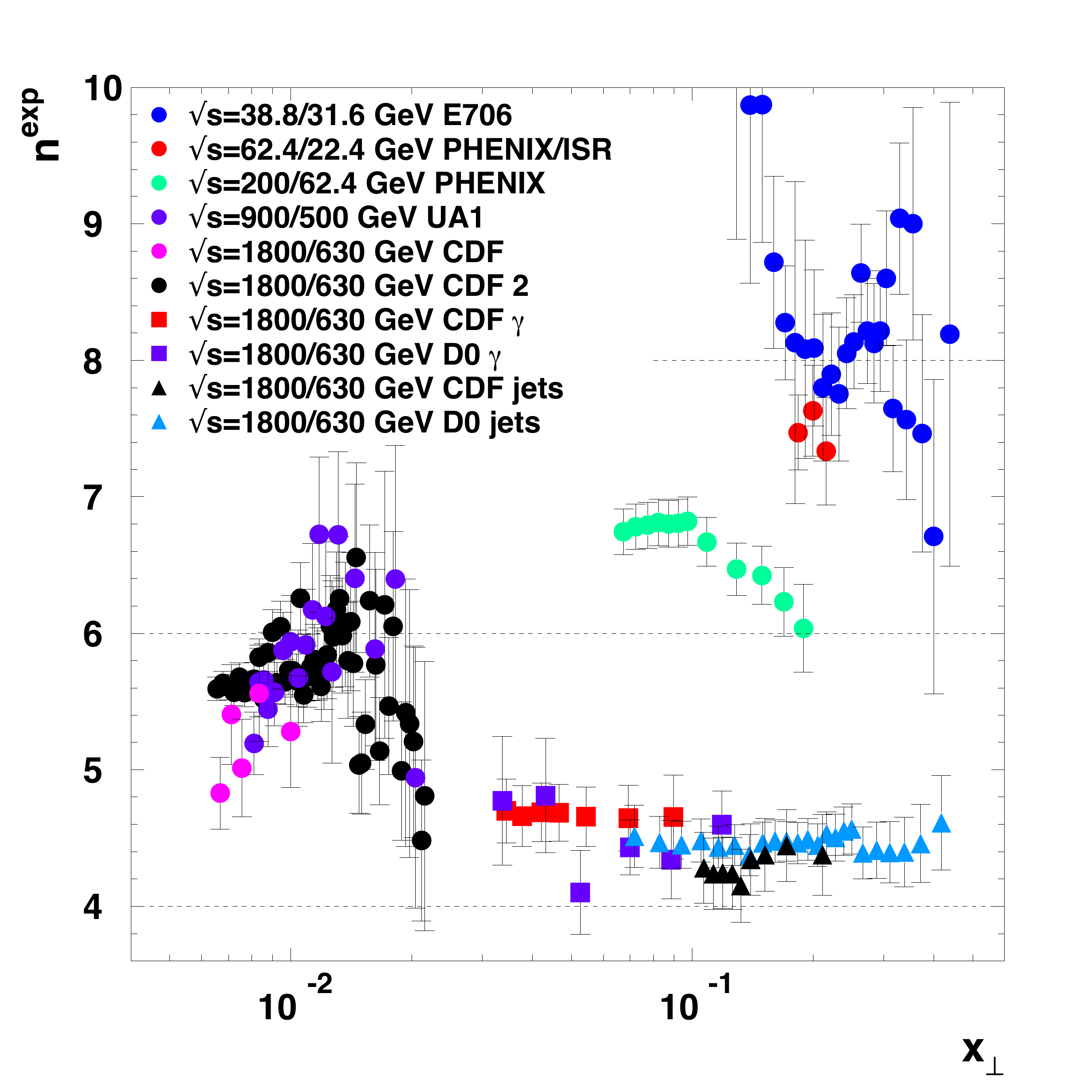}
\includegraphics[width=8cm]{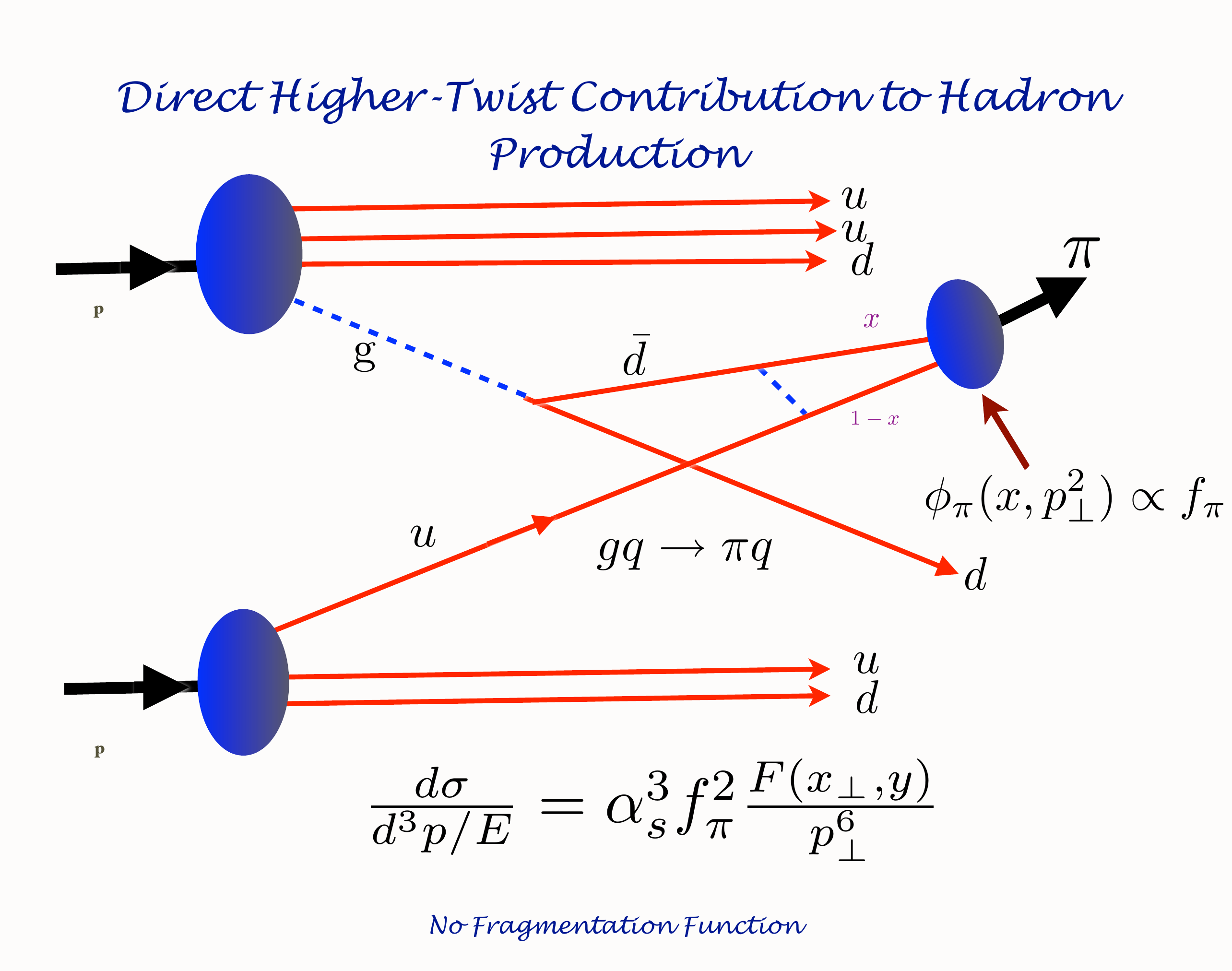}
\end{center}
\caption{ (a) Scaling of inclusive cross sections for hadron, photons and jets at high $p_T$ at fixed
$x_T= 2{p_T\over \sqrt s}.$ (b) Example of a direct QCD contribution for pion
production. ~\cite{Arleo:2009ch,Arleo:2010yg}.}
\label{Hiptscaling}
\end{figure}

The usual expectation is that leading-twist subprocesses (i.e, the leading power-law contributions) will dominate measurements of high $p_T$ hadron production at RHIC and  at Tevatron energies. In fact, the  data for isolated photon production $ p p \to \gamma_{\rm direct} X,$ as well as jet production, agrees well with the  leading-twist scaling prediction $n_{\rm eff}  \simeq 4.5$.~\cite{Arleo:2009ch}
However,   measurements  of  $n_{\rm eff} $ for hadron production  are not consistent with the leading twist predictions.
See Fig. \ref{Hiptscaling}(a). Striking
deviations from the leading-twist predictions were also observed at lower energy at the ISR and  Fermilab fixed-target experiments.~\cite{Sivers:1975dg} This deviation points to a significant contribution from direct higher-twist processes where the hadron is created directly in the hard subprocess rather than from quark or gluon jet fragmentation.

In fact, a significant fraction of high $p^H_\perp$ isolated hadrons can emerge
directly from hard higher-twist subprocess~\cite{Arleo:2009ch,Arleo:2010yg} even at the LHC.
An example is shown in  Fig. \ref{Hiptscaling}(b).  The direct production of hadrons can also explain~\cite{Brodsky:2008qp} the remarkable ``baryon anomaly" observed at RHIC:  the ratio of baryons to mesons at high $p^H_\perp$,  as well as the power-law fall-off $1/ p_\perp^n$ at fixed $x_\perp = 2 p_\perp/\sqrt s, $ both  increase with centrality,~\cite{Adler:2003kg} opposite to the usual expectation that protons should suffer more energy loss in the nuclear medium than mesons.
The high values $n_{\rm eff}$ with $x_T$ seen in the data  indicate the presence of an array of higher-twist processes, including subprocesses where the hadron enters directly,
rather than through jet fragmentation.~\cite{Blankenbecler:1972zz}   Although they are suppressed by powers of $1/p_T$, the direct  higher twist processes can dominate because they are energy efficient -- no same side energy or momentum is lost from the undetected fragments. Thus the incident colliding partons are evaluated at the minimum possible values of light-front momentum fractions $x_1$ and $x_2$, where the parton distribution functions are numerically large.


Normally many more pions than protons are produced at high transverse momentum in hadron-hadron collisions. This is also true for the peripheral collisions of heavy ions. However, when the nuclei collide with maximal overlap (central collisions) the situation is reversed -- more protons than pions emerge.  This observation at RHIC~\cite{Adler:2003kg}  contradicts the usual expectation that protons should be more strongly absorbed than pions in the nuclear medium.
This deviation also points to a significant contribution from direct higher twist processes where hadrons, particularly baryons are created directly in the hard subprocess rather than from quark or gluon jet fragmentation.
Since these processes  create color-transparent baryons, this mechanism can explain
the RHIC baryon anomaly.~\cite{Brodsky:2008qp}. Evidence for color transparency~\cite{Brodsky:1988xz} is particularly clear in diffractive dijet production on nuclei~\cite{Aitala:2000hc}

\section{\bf Breakdown of Perturbative QCD Factorization Theorems}

The factorization picture derived from the parton and pQCD has played a guiding role in virtually all aspects of hadron physics phenomenology.  In the case of inclusive reactions such as $ {E_H d\sigma\over d^3 p_H}( p p \to HX) $,  the pQCD ansatz predicts that the cross section at leading order in the transverse momentum  $p_T$ can be computed by convoluting the perturbatively calculable hard subprocess quark and gluon cross section with the process-independent structure functions of the colliding hadrons with the quark fragmentation functions.  The resulting cross section scales as $1/ p^4_T,$ modulo the DGLAP scaling violations derived from the logarithmic evolution of the structure functions and fragmentation distributions, as well as  the running of the QCD coupling appearing in the hard scattering subprocess matrix element.

The effects of final-state interactions of the scattered quark  in deep inelastic scattering  have been traditionally assumed to either give  an inconsequential phase factor or power-law suppressed corrections.  However, this is only true for sufficiently inclusive cross sections.  For example, consider semi-inclusive deep inelastic lepton scattering (SIDIS) on a polarized target $\ell p_\updownarrow \to H \ell' X.$  In this case the final-state gluonic interactions of the scattered quark lead to a  $T$-odd non-zero spin correlation of the plane of the lepton-quark scattering plane with the polarization of the target proton~\cite{Brodsky:2002cx} which is not power-law suppressed with increasing virtuality of the photon $Q^2$; i.e. it Bjorken-scales.    This  leading-twist  ``Sivers effect"~\cite{Sivers:1989cc} is nonuniversal in the sense that pQCD predicts an opposite-sign correlation in Drell-Yan reactions relative to single-inclusive deep inelastic scattering.~\cite{Collins:2002kn,Brodsky:2002rv}
This important but  yet untested prediction occurs because the Sivers effect in the Drell-Yan reaction is modified by  the initial-state interactions of the annihilating antiquark.

Similarly, the final-state interactions of the produced quark with its comoving spectators in SIDIS produces a final-state $T$-odd polarization correlation -- the ``Collins effect".  This can be measured without beam polarization by measuring the correlation of the polarization of a hadron such as the $\Lambda$ baryon with the quark-jet production plane.  Analogous spin effects occur in QED reactions due to the rescattering via final-state Coulomb interactions. Although the Coulomb phase for a given partial wave is infinite, the interference of Coulomb phases arising from different partial waves leads to observable effects.  These considerations have led to a reappraisal of the range of validity of the standard factorization ansatz.~\cite{Collins:2007nk}

\begin{figure}
 \begin{center}
\includegraphics[width=16cm]{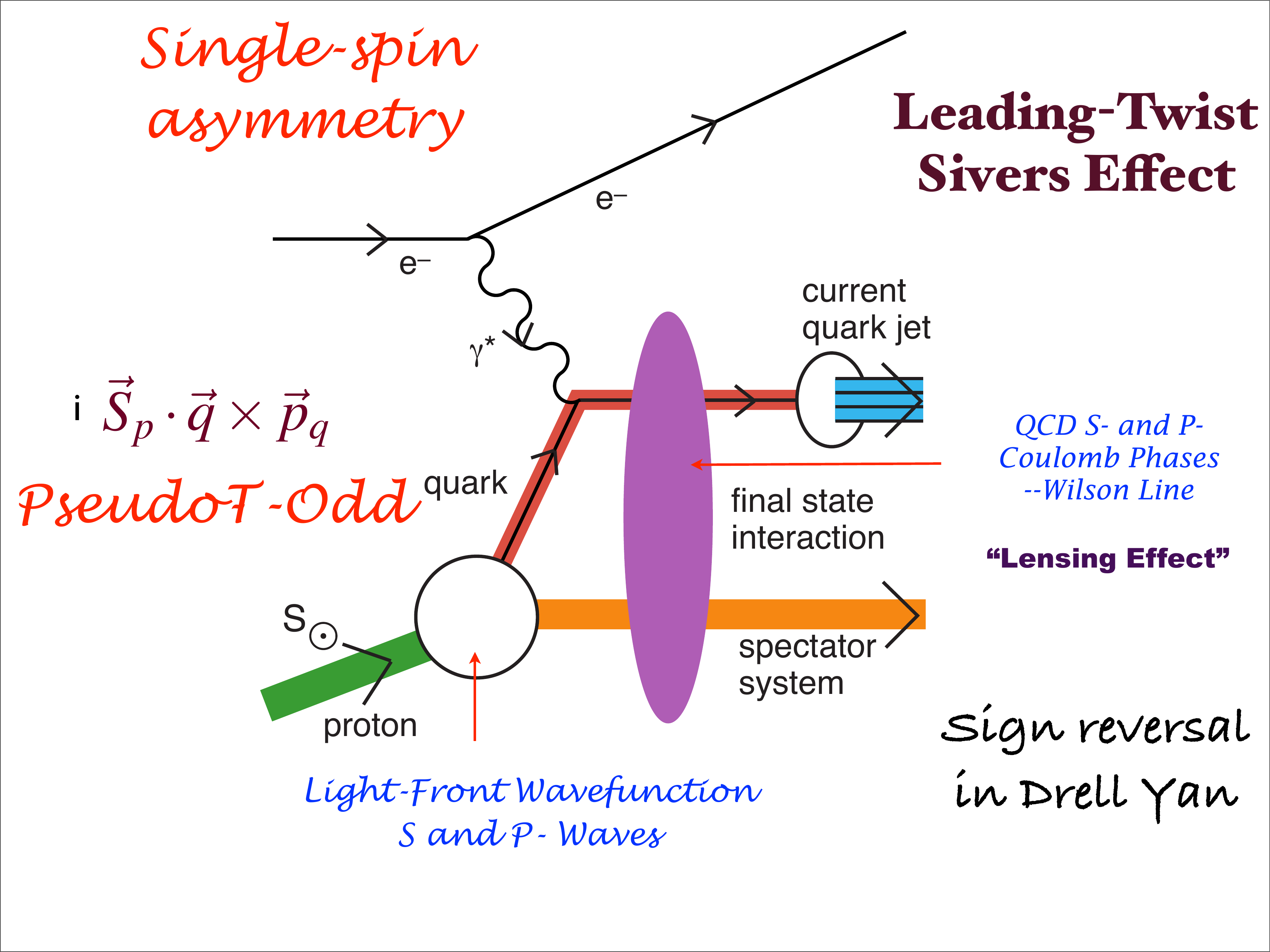}
\end{center}
\caption{Origin of the Sivers single-spin asymmetry in deep inelastic lepton scattering.}
\label{Sivers}
\end{figure}

The calculation of the Sivers single-spin asymmetry in deep inelastic lepton scattering in QCD is illustrated in
Fig.~\ref{Sivers}.
The analysis requires two different orbital angular momentum components: $S$-wave with the quark-spin parallel to the proton spin and $P$-wave for the quark with anti-parallel spin; the difference between the final-state ``Coulomb" phases leads to a $\vec S \cdot \vec q \times \vec p$ correlation of the proton's spin with the virtual photon-to-quark production plane.~\cite{Brodsky:2002cx}  Thus, as it is clear from its QED analog,  the final-state gluonic interactions of the scattered quark lead to a  $T$-odd non-zero spin correlation of the plane of the lepton-quark scattering plane with the polarization of the target proton.~\cite{Brodsky:2002cx}

The  $S$- and $P$-wave proton wavefunctions also appear in the calculation of the Pauli form factor quark-by-quark. Thus one can correlate the Sivers asymmetry for each struck quark with the anomalous magnetic moment of the proton carried by that quark,~\cite{Lu:2006kt}  leading to the prediction that the Sivers effect is larger for  positive pions as seen by the
HERMES experiment at DESY,~\cite{Airapetian:2004tw} the COMPASS experiment~\cite{Bradamante:2011xu,Alekseev:2010rw,Bradamante:2009zz} at CERN, and CLAS at Jefferson Laboratory~\cite{Avakian:2010ae,Gao:2010av}

This  leading-twist Bjorken-scaling ``Sivers effect"  is nonuniversal since QCD predicts an opposite-sign correlation~\cite{Collins:2002kn,Brodsky:2002rv} in Drell-Yan reactions due to the initial-state interactions of the annihilating antiquark.
The  $S-$ and $P$-wave proton wavefunctions also appear in the calculation of the Pauli form factor quark-by-quark. Thus one can correlate the Sivers asymmetry for each struck quark with the anomalous magnetic moment of the proton carried by that quark~\cite{Lu:2006kt},  leading to the prediction that the Sivers effect is larger for  positive pions.

The physics of the ``lensing dynamics" or Wilson-line physics~\cite{Brodsky:2010vs} underlying the Sivers effect  involves nonperturbative quark-quark interactions at small momentum transfer, not  the hard scale $Q^2$  of the virtuality of the photon.  It would interesting to see if the strength of the
soft initial- or final- state scattering can be predicted using the effective confining potential of QCD from light-front holographic QCD.

Measurements~\cite{Falciano:1986wk} of the Drell-Yan Process $\pi p \to \mu^+ \mu^- X$ display an angular distribution which contradicts pQCD expectations. In particular one observes an anomalously large $\cos{ 2\phi} $ azimuthal angular correlation between the lepton decay plane and its production plane which contradicts the Lam-Tung relation, a prediction of perturbative QCD factorization.~\cite{Lam:1980uc}    Such effects again point to the importance of initial and final-state interactions of the hard-scattering constituents,~\cite{Boer:2002ju}  corrections not included in the standard pQCD factorization formalism.

As  noted by Collins and Qiu,~\cite{Collins:2007nk} the traditional factorization formalism of perturbative QCD  fails in detail for many hard inclusive reactions because of initial- and final-state interactions.  For example, if both the
quark and antiquark in the Drell-Yan subprocess
$q \bar q \to  \mu^+ \mu^-$ interact with the spectators of the
other  hadron, then one predicts a $\cos 2\phi \sin^2 \theta$ planar correlation in unpolarized Drell-Yan
reactions.~\cite{Boer:2002ju}  This ``double Boer-Mulders effect" can account for the anomalously large $\cos 2 \phi$ correlation and the corresponding violation~\cite{Boer:2002ju, Boer:1999mm} of the Lam Tung relation for Drell-Yan processes observed by the NA10 collaboration.~\cite{Falciano:1986wk}    Such effects again point to the importance of initial and final-state interactions of the hard-scattering constituents, corrections not included in the standard pQCD factorization formalism.
One also observes large single spin asymmetries in reactions such as $ p p_\updownarrow \pi X$, an effect not yet explained.~\cite{Liang:1993rz} Another important signal for factorization breakdown at the LHC  will be the observation of a $\cos 2 \phi$ planar correlation in dijet production.

The final-state interactions of the struck quark with the target spectators~\cite{Brodsky:2002ue}  also lead to diffractive events in deep inelastic scattering (DDIS) at leading twist,  such as $\ell p \to \ell' p' X ,$ where the proton remains intact and isolated in rapidity;    in fact, approximately 10 \% of the deep inelastic lepton-proton scattering events observed at HERA are
diffractive.~\cite{Adloff:1997sc, Breitweg:1998gc} This seems surprising since the underlying hard subprocess $\ell q \to \ell^\prime q^\prime$ is highly disruptive of the target nucleon.
The presence of a rapidity gap
between the target and diffractive system requires that the target
remnant emerges in a color-singlet state; this is made possible in
any gauge by the soft rescattering incorporated in the Wilson line or by augmented light-front wavefunctions.
Quite different fractions of single $ p p \to {\rm Jet}  \,p^\prime  X$ and double diffractive
$p \bar p \to {\rm Jet} \, p^\prime  \bar p^\prime X$ events are observed at the Tevatron. The underlying mechanism is believed to be soft gluon exchange between the scattered quark and the remnant system in the final state occurring after the hard scattering  occurs.

One can show~\cite{Stodolsky:1994ka} using Gribov-Glauber theory  that the Bjorken-scaling diffractive deep inelastic scattering events lead to the shadowing of nuclear structure functions at small $x_{\rm Bjorken}.$  This is due to the destructive interference of two-step and one step amplitudes in the nucleus.   Since diffraction involves rescattering, one sees that shadowing and diffractive processes  are not intrinsic properties of hadron and nuclear wavefunctions and structure functions, but are properties of the complete dynamics of the scattering reaction.~\cite{Brodsky:2008xe}


The CDF~\cite{Aaltonen:2011kc} and D0~\cite{:2007qb} experiments at the Tevatron have recently reported that the $t$ and  $\bar t$ heavy quarks do not have the same momentum distributions in $ \bar p  p \to t \bar t X$ events.  The observed asymmetry is much larger than predicted from QCD NLO corrections to the $\bar q q \to t \bar t$ subprocess. The Tevatron $t~\bar t$ asymmetry may indicate the importance of rescattering Coulomb-like final state interactions of the top quarks with $ud$ and $\bar u \bar d$ remnant spectators of the colliding proton and antiproton.~\cite{BVHZ}.  This effect can also lead to a $t ~\bar t$  asymmetry  in $p p \to t \bar t  X$ collisions at the LHC  since the $t$ quark can be color-attracted of one of the spectator $ud$ diquarks produced in the $q \bar q \to t \bar t$ subprocess; however, the effect would only significant when the $t$ and $ud$ systems have small rapidity separation.~\cite{BVHZ}

\section{\bf Non-Universal Antishadowing}

\begin{figure}
 \begin{center}
\includegraphics[width=16cm]{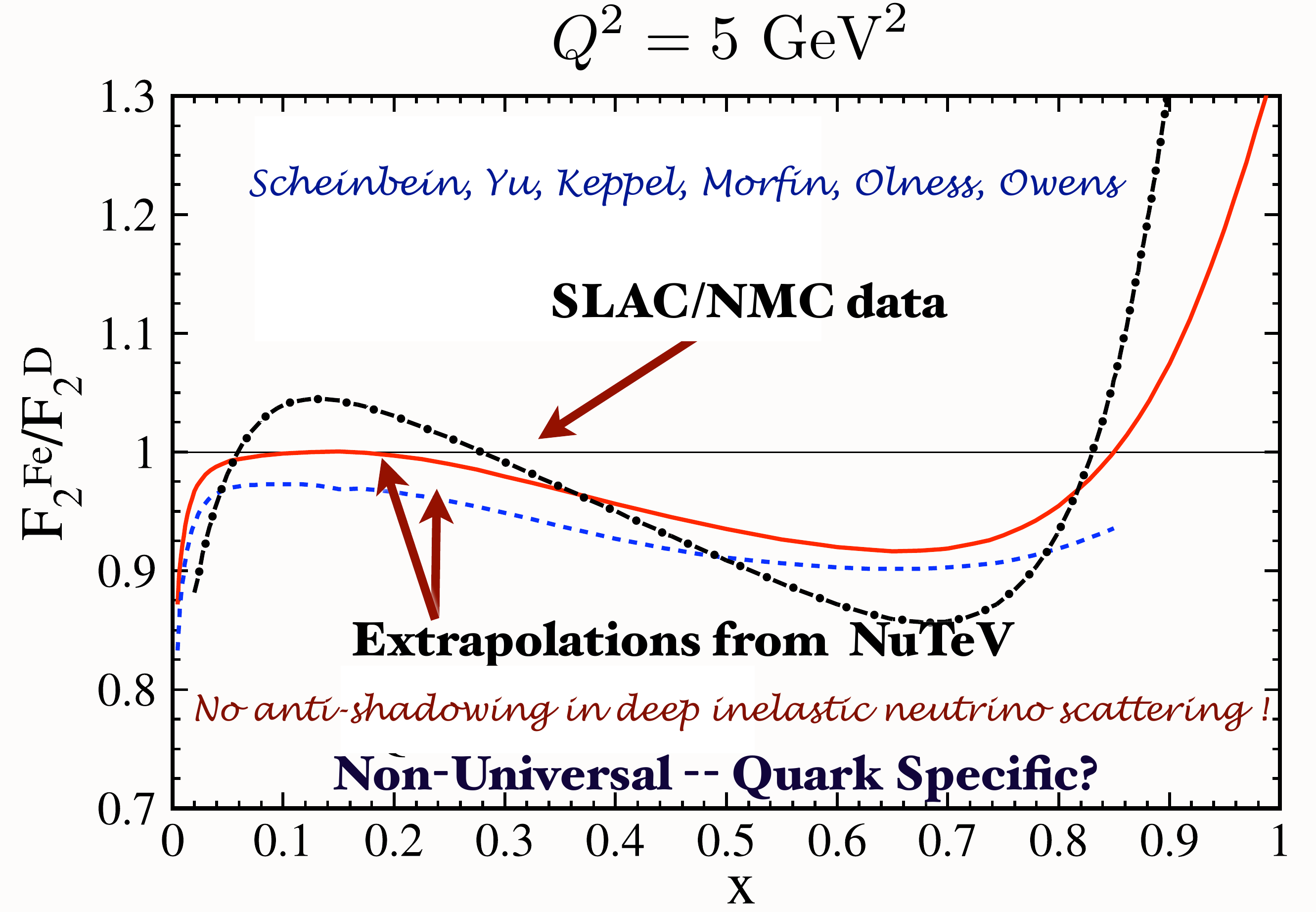}
\end{center}
\caption{Antishadowing is nonuniversal.}
\label{Antishadowing}
\end{figure}

It has been conventional to assume that the nuclear modifications to the structure functions measured in deep inelastic charged lepton-nucleus and neutrino-nucleus interactions are identical.  In fact,  Gribov-Glauber theory predicts that the antishadowing of nuclear structure functions is not  universal, but depends on the quantum numbers of each struck quark and antiquark.~\cite{Brodsky:2004qa}  This observation can explain the recent analysis of Schienbein {\it et al.}~\cite{Schienbein:2008ay} which finds that the NuTeV measurements of nuclear structure functions  obtained from neutrino  charged current reactions differ significantly from the distributions measured in deep inelastic electron and muon scattering.  See Fig. \ref{Antishadowing}.  This implies that part of
of the anomalous NuTeV result for $\theta_W$ could be due to the non-universality of nuclear antishadowing for charged and
neutral currents.

The {\it antishadowing} of the nuclear structure functions as observed in deep
inelastic lepton-nucleus scattering is particularly interesting. Empirically, one finds $R_A(x,Q^2) \equiv\\  \left(F_{2A}(x,Q^2)/ (A/2) F_{d}(x,Q^2)\right)
> 1 $ in the domain $0.1 < x < 0.2$;  { i.e.}, the measured nuclear structure function (referenced to the deuteron) is larger than the
scattering on a set of $A$ independent nucleons.
There are leading-twist diffractive contributions $\gamma^* N_1 \to (q \bar q) N_1$  arising from Reggeon exchanges in the
$t$-channel. For example, isospin--non-singlet $C=+$ Reggeons contribute to the difference of proton and neutron
structure functions, giving the characteristic Kuti-Weiskopf $F_{2p} - F_{2n} \sim x^{1-\alpha_R(0)} \sim x^{0.5}$ behavior at small $x$. The
$x$ dependence of the structure functions reflects the Regge behavior $\nu^{\alpha_R(0)} $ of the virtual Compton amplitude at fixed $Q^2$ and
$t=0.$ The phase of the diffractive amplitude is determined by analyticity and crossing to be proportional to $-1+ i$ for $\alpha_R=0.5,$ which
together with the phase from the Glauber cut, leads to {\it constructive} interference of the diffractive and nondiffractive multi-step nuclear
amplitudes.  The nuclear structure function is predicted~ \cite{Brodsky:1989qz} to be enhanced precisely in the domain $0.1 < x <0.2$ where
antishadowing is empirically observed.  The strength of the Reggeon amplitudes is fixed by the fits to the nucleon structure functions, so there
is little model dependence.
Since quarks of different flavors couple to different Reggeons, this leads to the remarkable prediction that
nuclear antishadowing is not universal;~\cite{Brodsky:2004qa}  it depends on the quantum numbers of the struck quark. This picture implies substantially different
antishadowing for charged and neutral current reactions, thus affecting the extraction of the weak-mixing angle $\theta_W$.

\section{Dynamic versus Static Hadronic Structure Functions}
The nontrivial effects from rescattering and diffraction highlight the need for a fundamental understanding the dynamics of hadrons in QCD at the amplitude
level. This is essential for understanding phenomena such as hadronization; i.e.,  the quantum mechanics of hadron formation, the remarkable
effects of initial and final interactions, the origins of diffractive phenomena and single-spin asymmetries, and manifestations of higher-twist
semi-exclusive hadron subprocesses.

It is usually assumed --  following the intuition of the parton model -- that the structure functions  measured in deep inelastic scattering can be computed in the Bjorken-scaling leading-twist limit from the absolute square of the light-front wavefunctions, summed over all Fock states.  In fact,  dynamical effects, such as the Sivers spin correlation and diffractive deep inelastic lepton scattering due to final-state gluon interactions,  contribute to the experimentally observed deep inelastic lepton-hadron cross sections.
Diffractive events also lead to the interference of two-step and one-step processes in nuclei which in turn, via the Gribov-Glauber theory, lead to the shadowing and the antishadowing of the deep inelastic nuclear structure functions;~\cite{Brodsky:2004qa}  such lensing phenomena are not included in the light-front wavefunctions of the nuclear eigenstate.
This leads to an important  distinction between ``dynamical''  vs. ``static''  (wavefunction-specific)
structure functions.~\cite{Brodsky:2009dv}

\begin{figure}
 \begin{center}
\includegraphics[width=14cm]{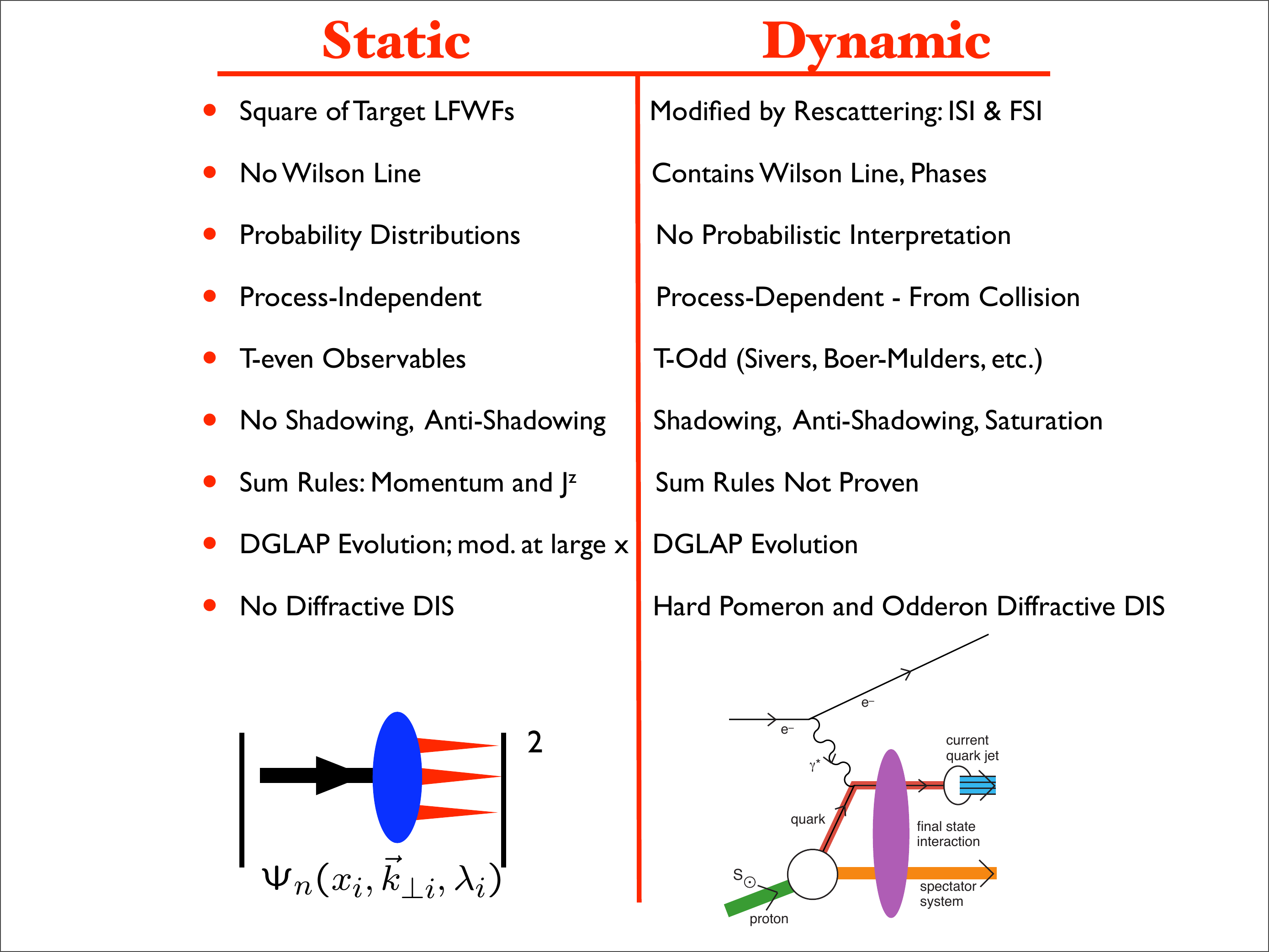}
\end{center}
\caption{Static versus dynamic structure functions.}
\label{figNew17}
\end{figure}

It is thus important to distinguish~\cite{Brodsky:2009dv} ``static" structure functions which are computed directly from the light-front wavefunctions of  a target hadron from the nonuniversal ``dynamic" empirical structure functions which take into account rescattering of the struck quark in deep inelastic lepton scattering.
See  Fig. \ref{figNew17}.
The real wavefunctions of hadrons which underly the static structure functions cannot describe diffractive deep inelastic scattering nor  single-spin asymmetries, since such phenomena involve the complex phase structure of the $\gamma^* p $ amplitude.
One can augment the light-front wavefunctions with a gauge link corresponding to an external field
created by the virtual photon $q \bar q$ pair
current,~\cite{Belitsky:2002sm,Collins:2004nx} but such a gauge link is
process dependent, so the resulting augmented
wavefunctions are not universal.~\cite{Collins:2002kn} The physics of rescattering and nuclear shadowing is not
included in the nuclear light-front wavefunctions and a
probabilistic interpretation of the nuclear DIS cross section in terms of hadron structure is thus
precluded in principle, although it can often be treated as an effective approximation.

\section{\bf The Principle of Maximum Conformality and the Elimination of the Renormalization Scale Ambiguity}

A key difficulty in making precise perturbative QCD predictions is
the uncertainty in determining the renormalization scale $\mu$ of the
running coupling $\alpha_s(\mu^2)$. It is common practice to simply guess
a physical scale $\mu = Q$ of order of a typical momentum
transfer $Q$ in the process, and then vary the scale over a range
$Q/2$ and $2 Q$.  This procedure is clearly problematic, since the
resulting fixed-order pQCD prediction will depend on the choice of
renormalization scheme;  it can even predict negative QCD cross
sections at next-to-leading-order.  If one uses the  criterion that one should choose the renormalization scale to have minimum sensitivity, one gets the wrong answer in QED and even in QCD. The prediction violates the transitivity property of the renormalization group; it will also depend on the choice of renormalization scheme.  Worse,  if one tries to minimize sensitivity, the resulting renormalization scale goes to zero as the gluon jet virtuality becomes large in $e^+ e^- \to q \bar q g$ three-jet events.~\cite{Kramer:1987dd}

The  running
coupling in any gauge theory  sums  all terms involving the
$\beta$ function; in fact, when the renormalization scale is set
properly, all non-conformal $\beta \ne 0$ terms  in a perturbative
expansion arising from renormalization are summed into the running
coupling.
The remaining terms in the perturbative series are then
identical to those of a conformal theory; i.e., the corresponding
theory with $\beta=0$.
As discussed by Di Giustino, Wu, and myself~\cite{Brodsky:2011ig,Brodsky:2011ta}, the resulting scale-fixed predictions using
this  ``principle of maximum conformality"  are independent of
the choice of renormalization scheme --  a key requirement of
renormalization group invariance.     In practice,  the scale can often be determined from the $n_f$ dependence of the NLO terms.  The BLM/PMC scale also determines the number of effective flavors in the $\beta$-function. The results avoid renormalon
resummation and agree with QED scale-setting in the Abelian limit.
The PMC is the principle~\cite{Brodsky:2011ig,Brodsky:2011ta}
which underlies the BLM scale-setting
method.~\cite{Brodsky:1982gc}

Extended renormalization group equations, which express the invariance of physical observables under both the renormalization scale- and scheme-parameter transformations, provide a convenient way for analyzing the scale- and scheme-dependence of the physical process. In a recent paper~\cite{Brodsky:2011ta} , we have analyzed the scale-dependence of the extended renormalization group equations at the four-loop level. Using the principle of maximum conformality, all non-conformal $\{\beta_i\}$ terms in the perturbative expansion series can be summed into the running coupling, and the resulting scale-fixed predictions are verified to be independent of the renormalization scheme. Different schemes lead to different effective PMC/BLM scales, but the final results are scheme independent. Conversely, from the requirement of scheme independence, one not only can obtain scheme-independent commensurate scale relations among different observables, but also determine the scale displacements among the PMC/BLM scales which are derived under different schemes. In principle, the PMC/BLM scales can be fixed order-by-order, and as a useful reference, we present a systematic and scheme-independent procedure for setting PMC/BLM scales up to NNLO.

Thus, most important, the BLM/PMC method gives results which are independent of the choice of renormalization scheme at each order of perturbation theory, as required by the transitivity property of the renormalization group. The argument of the running coupling constant acquires the appropriate displacement appropriate to its scheme so that the evaluated result is scheme-independent. In the case of Abelian theory, the scale is proportional to the photon virtuality and sums all vacuum polarization corrections to all orders.

The elimination
of the renormalization scheme ambiguity will not
only increase the precision of QCD tests,  but it will also
increase the sensitivity of LHC experiments and other measurements to new physics
beyond the Standard Model. The BLM/PMC method also provides scale-fixed,
scheme-independent high-precision connections between observables, such as the ``Generalized Crewther Relation'',~\cite{Brodsky:1995tb} as well as other ``Commensurate Scale Relations''.~\cite{Brodsky:1994eh,Brodsky:2000cr}  Clearly the elimination of the renormalization scale ambiguity would greatly improve the precision of QCD predictions and increase the sensitivity of searches for  new physics at the LHC.

 \section{Light-Front Quantization}

The distributions of electrons within an atom are determined in QED using the Schr\"odinger wavefunction, the eigenfunction of the QED Hamiltonian.  In principle, one could  calculate hadronic spectroscopy and wavefunctions by solving for the eigenstates of the QCD Hamiltonian:
$H \vert  \Psi \rangle = E \vert \Psi \rangle$
 at fixed time $t.$ However, this traditional method -- called the ``instant" form" by Dirac,~\cite{Dirac:1949cp} is  plagued by complex vacuum and relativistic effects, as well
as  by  the fact that the boost of such fixed-$t$ wavefunctions away from the hadron's rest frame is an intractable dynamical problem.
However, there is an extraordinarily powerful non-perturbative alternative -- quantization at fixed light-front (LF) time $\tau = t + z/c = x^+ = x^0 + x^3$ -- the ``front-form" of Dirac.~\cite{Dirac:1949cp} In this framework each hadron $H$ is identified as an eigenstate of the QCD Hamiltonian
$H_{LF}^{QCD} \vert \Psi_H \rangle = M^2_H \vert \Psi_H \rangle$,
where  $H_{LF}^{QCD} = P_\mu P^\mu= P^- P^+ -  P^2_\perp$ is derived directly from the QCD Lagrangian or action. The eigenvalues of this Heisenberg equation give the complete mass spectrum of hadrons. The eigensolution  $|\Psi_H \rangle$  projected on the free Fock basis  provides  the  set of valence and non-valence  light-front Fock state wavefunctions $\Psi_{n/H}(x_i, k_{\perp i}, \lambda_i)$, which describe the hadron's momentum and spin distributions and  the direct measures of its structure at the quark and gluon level.
If one quantizes the gluon field in light-cone gauge $A^+= A^0 + A^3=0$, the gluons have physical polarization $S^z = \pm 1$, there are no ghosts, so that  one has a physical interpretation of the quark and gluon constituents.
The constituents of a bound state in a light-front wavefunction are measured at the same light-front time $\tau$ -- along the front of a light-wave, as in a flash picture.  In contrast, the constituents of a bound state in an instant form wavefunction must be measured at the same instant time $t$ -  - this requires the exact synchrony in time of many simultaneous probes.

A remarkable feature of LFWFs is the fact that they are frame
independent; i.e., the form of the LFWF is independent of the
hadron's total momentum $P^+ = P^0 + P^3$ and $P_\perp.$
The boost invariance of  LFWFs contrasts dramatically with the complexity of  boosting the wavefunctions defined at fixed time $t.$~\cite{Brodsky:1968ea}
Light-front quantization is thus the ideal framework to describe the
structure of hadrons in terms of their quark and gluon degrees of freedom.  The
constituent spin and orbital angular momentum properties of the
hadrons are also encoded in the LFWFs.
The total  angular momentum projection~\cite{Brodsky:2000ii}
$J^z = \sum_{i=1}^n  S^z_i + \sum_{i=1}^{n-1} L^z_i$
is conserved Fock-state by Fock-state and by every interaction in the LF Hamiltonian.
The constituent spin and orbital angular momentum properties of the hadrons are thus encoded in their LFWFs.
The empirical observation that quarks carry only a small fraction of the nucleon angular momentum highlights the importance of quark orbital angular momentum.  In fact the nucleon anomalous moment and the Pauli form factor are zero unless the quarks carry nonzero $L^z$.

Hadron observables, e.g., hadronic structure functions, form factors, distribution amplitudes,  GPDs, TMDs, and Wigner distributions can be computed as simple convolutions of light-front wavefunctions (LFWFs).  For example,  one can calculate the electromagnetic and gravitational form factors
$<p+ q| j^\mu(0)| p>$ and $<p+ q| t^{\mu \nu}(0)| p>$ of a hadron from the Drell-Yan-West formula -- i.e., the overlap of LFWFs.
The anomalous gravitomagnetic moment $B(0)$ defined from the spin-flip matrix element
$<p+ q| t^{\mu \nu}(0)| p>$ at $ q\to 0$ vanishes -- consistent with the equivalence theorem of gravity.
In contrast, in the instant form, the overlap of instant time wavefunctions is not sufficient.  One must also couple the photon probe to currents arising spontaneously from the vacuum which are connected to the hadron's constituents.
The Light-Front method is  directly applicable for describing atomic bound states in both the relativistic and nonrelativistic domains; it is particularly useful for atoms in flight since the LFWFs are frame-independent. It also satisfies theorems
such as cluster decomposition.

One can solve the LF Hamiltonian problem for theories  in one-space and one-time  by Heisenberg matrix diagonalization. For example, the complete set of discrete and continuum eigensolutions of mesons and baryons  in QCD(1+1) can be obtained to any desired precision for general color,  multiple flavors, and general quark masses using the discretized light-cone quantized (DLCQ) method.~\cite{Pauli:1985ps,Hornbostel:1988fb}  The  DLCQ approach can in principle be applied to QED(3+1) and QCD(3+1); however,  in practice, the  huge matrix diagonalization problem is computational challenging.

\section{AdS/QCD and Light-Front Holography \label{LFHsec}}

A long-sought goal in hadron physics is to find a simple analytic first approximation to QCD analogous
to the Schr\"odinger-Coulomb equation of atomic physics.	This problem is particularly challenging since the formalism must be relativistic, color-confining, and consistent with chiral symmetry.
de T\'eramond and I~\cite{deTeramond:2008ht} have shown that  the gauge/gravity duality leads  to a simple analytical
 and phenomenologically compelling nonperturbative approximation to the full light-front  QCD Hamiltonian -- ``Light-Front Holography".~\cite{deTeramond:2008ht}  Light-Front Holography  is in fact one of the most remarkable features of
 the AdS/CFT correspondence.~\cite{Maldacena:1997re}  In particular
 the soft-wall AdS/QCD model, modified by a positive-sign dilaton metric, leads to a simple
Schr\"odinger-like light-front wave equation and a remarkable one-parameter description of nonperturbative hadron dynamics~\cite{deTeramond:2008ht,deTeramond:2009xk}. The model predicts a zero-mass pion for massless quarks and a Regge spectrum of linear trajectories with the same slope in the (leading) orbital angular momentum $L$ of the hadrons and their radial  quantum number $N$.

Light front holographic methods
allow one to project the functional dependence of the wavefunction $\Phi(z)$ computed  in the  AdS fifth dimension to the  hadronic frame-independent light-front wavefunction $\psi(x_i, b_{\perp i})$ in $3+1$ physical space-time. The variable $z $ maps  to a transverse
LF variable $ \zeta(x_i, b_{\perp i})$.
The result is a single-variable light-front Schr\"odinger equation which determines the eigenspectrum and the LFWFs of hadrons for general spin and orbital angular momentum.  The transverse coordinate $\zeta$ is closely related to the invariant mass squared  of the constituents in the LFWF  and its off-shellness  in  the LF kinetic energy,  and it is thus the natural variable to characterize the hadronic wavefunction.  In fact $\zeta$ is the only variable to appear
in the relativistic light-front Schr\"odinger equations predicted from
holographic QCD  in the limit of zero quark masses.
The coordinate $z$ in AdS space is thus uniquely identified with  a Lorentz-invariant  coordinate $\zeta$ which measures the separation of the constituents within a hadron at equal light-front time.

The result is
a semi-classical frame-independent first approximation to the spectra and light-front wavefunctions of meson and baryon light-quark  bound states,  which in turn predicts  the
behavior of the pion and nucleon  form factors.
The hadron eigenstates generally have components with different orbital angular momentum; e.g.,  the proton eigenstate in AdS/QCD with massless quarks has $L^z=0$ and $L^z=1$ light-front Fock components with equal probability.
Thus in AdS/QCD the spin of the proton is carried by the quark orbital angular momentum: $J^z=  \langle L^z\rangle=\pm 1/2$ since $\langle\sum S^z_q \rangle= 0,$~\cite{Brodsky:2011zj}  helping to explain the ``spin-crisis".

\begin{figure}
 \begin{center}
\includegraphics[width=18cm]{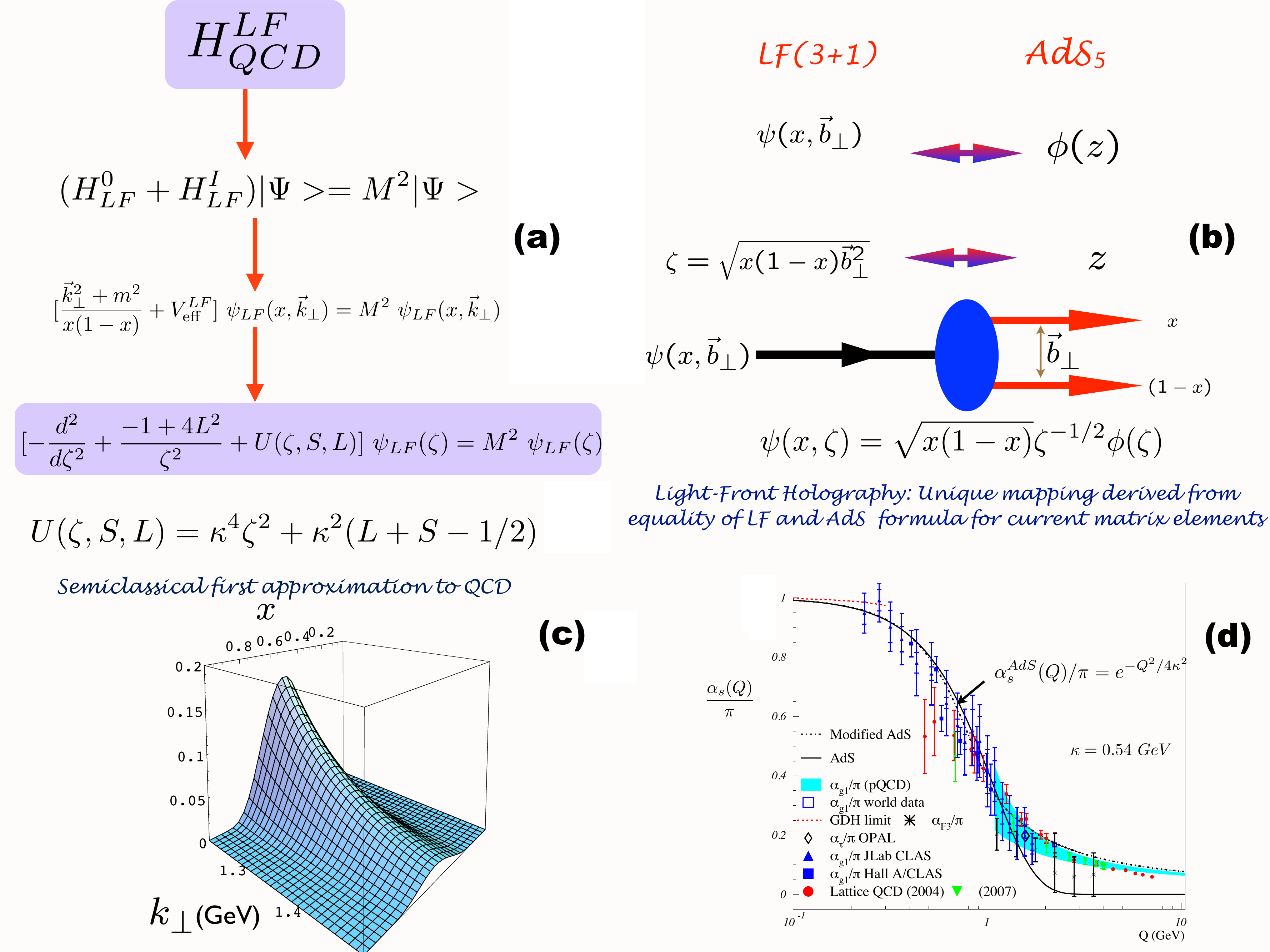}
\end{center}
\caption{(a) Reduction of the Light-Front Hamiltonian to an effective LF Schrodinger Equation for mesons.
(b) Mapping of the fifth dimension coordinate $z$ to the invariant LF separation variable $\zeta$.
The insert  (c) shows the AdS/QCD -- light-front holography prediction
for the pion's valence LFWF $\psi(x, \mbf{k}_\perp).$  (d) The running coupling predicted by AdS/QCD
normalized to $\alpha_s/\pi = 1$ compared with the effective charge defined from the Bjorken sum rule.
From Ref.~\cite{deTeramond:2008ht}.}
\label{HLF}
\end{figure}

The AdS/QCD soft-wall model also predicts the form of the non-perturbative effective coupling $\alpha_s^{AdS}(Q)$ as shown in fig. \ref{HLF}(d) and its $\beta$-function in excellent agreement with
JLAB measurements.~\cite{Brodsky:2010ur} The AdS/QCD light-front wavefunctions also lead to a proposal for computing the hadronization of quark and gluon jets at the amplitude level.~\cite{Brodsky:2008tk}

In general the QCD Light-Front Hamiltonian can be systematically reduced to an effective equation in  acting on the valence Fock state.
This is illustrated for mesons in fig.~\ref{HLF}
The kinetic energy contains a term $L^2/ \zeta^2$ analogous to $\ell(\ell+1)/r^2$ in nonrelativistic theory, where the invariant $\zeta^2 = x(1-x)b^2_\perp$  is conjugate to the $q \bar q $ invariant mass  $k^2_\perp/x(1-x)$. It  plays the role of the radial variable $r$.  Here $L= L^z$ is the projection of the orbital angular momentum appearing in the $\zeta, \phi$ basis. In QCD, the interaction $U$ couples  the valence state to all Fock states.  The AdS/QCD model has the identical structure as the reduced form of the LF Hamiltonian, but it also specifies the confining potential as $U(\zeta,S,L) = \kappa^4\zeta^2+ \kappa^2(L + S - 1/2).$ This correspondence, plus the fact that one can match the AdS/QCD formulae for elastic electromagnetic and gravitational form factors to the LF Drell-Yan West formula, is the basis for light-front holography.  The light-quark meson and baryon spectroscopy is  well described taking the mass parameter $\kappa \simeq 0.5$ GeV.  The linear trajectories in $M^2_H(n,L)$  have the same slope in $L$ and $n$, the radial quantum number.  The corresponding LF wavefunctions are functions of the off-shell invariant mass.   AdS/QCD, together with Light-Front Holography~\cite{deTeramond:2008ht} thus provides a simple Lorentz-invariant color-confining approximation to QCD which is successful in accounting for light-quark meson and baryon spectroscopy as well as their LFWFs.
This semiclassical approximation to light-front QCD
is expected to break down at short distances
where hard gluon exchange and quantum corrections become important.
The model can be systematically improved by Lippmann-Schwinger methods~\cite{Chabysheva:2011fr} or using the AdS/QCD orthonormal basis to diagonalize the LF Hamiltonian.
One  can also
improve the semiclassical approximation by introducing nonzero quark masses and short-range Coulomb
corrections,  thus extending the predictions of the model to the dynamics and spectra of heavy and heavy-light quark systems.~\cite{Branz:2010ub}

\section {\bf QCD Condensates and the Cosmological Constant}

\begin{figure}
 \begin{center}
\includegraphics[width=16cm]{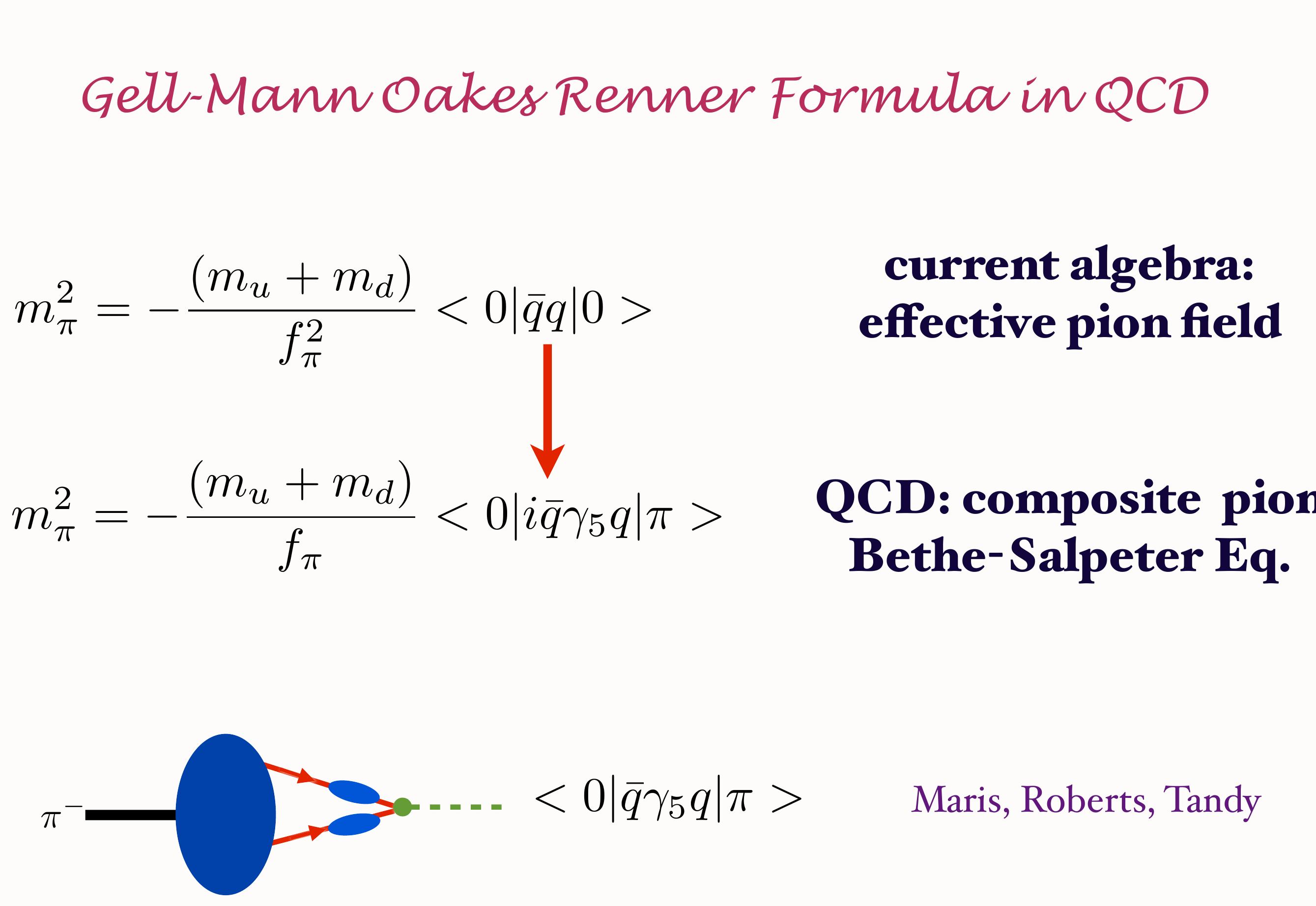}
\end{center}
\caption{Revised GMOR Relation.}
\label{GMOR}
\end{figure}

It is conventional to assume that the vacuum of QCD contains quark $\langle 0 \vert q \bar q \vert 0 \rangle$ and gluon  $\langle 0 \vert  G^{\mu \nu} G_{\mu \nu} \vert 0 \rangle$ vacuum condensates. However, as reviewed
by Zee~\cite{Zee:2008zz}, the resulting vacuum energy density from QCD leads to a $10^{45}$  order-of-magnitude or more discrepancy with the
measured cosmological constant.   In fact,  Zee has called this conflict ``one of the gravest puzzles of theoretical physics."
This  extraordinary contradiction between theory and cosmology has been used as an argument for the anthropic principle.~\cite{Weinberg:1987dv}
The  resolution of this long-standing puzzle has  been suggested~\cite{Brodsky:2008xu}, motivated by
Bethe-Salpeter and light-front analyses in which the QCD condensates are identified as ``in-hadron'' condensates, rather than  vacuum entities, but consistent with the Gell Mann-Oakes-Renner  relation.~\cite{Brodsky:2010xf}
See. Fig. \ref{GMOR}.
The ``in-hadron''  condensates become realized as higher Fock states of the hadron when the theory is quantized at fixed light-front time $\tau = t -z/c$.

Hadronic condensates have played an important role in quantum chromodynamics (QCD).
Conventionally, these condensates are considered to be properties
of the QCD vacuum and hence to be constant throughout space-time.
Recently a new perspective on the nature of QCD
condensates $\langle \bar q q \rangle$ and $\langle
G_{\mu\nu}G^{\mu\nu}\rangle$, particularly where they have spatial and temporal
support,
has been presented.~\cite{Brodsky:2009zd}
A  key ingredient in this approach is the use of Dirac's ``Front Form";~\cite{Dirac:1949cp} i.e.,  the light-front
(infinite momentum) frame to analyze the condensates.
In this formulation the spatial support of condensates
is restricted to the interior
of hadrons, since in the LF vacuum is an empty Fock state.  Thus
condensates
arise due to the interactions of quarks and
gluons which are confined within hadrons.

Physical eigenstates are built from operators acting on the vacuum.
It is thus important to distinguish two very different concepts of the vacuum in quantum field theories such as QED and QCD.   The conventional instant-form vacuum is a state defined at the same time $t$ at all spatial points in the universe.  In contrast, the front-form vacuum only senses phenomena which are causally connected; i.e., or within the observer's light-cone.
The instant-form vacuum is defined as the lowest energy eigenstate of the instant-form Hamiltonian.  For example, the instant-form vacuum in QED is saturated with quantum loops of leptons and photons. In calculations of physical processes one must then normal-order the vacuum and divide the $S$-matrix elements by the disconnected vacuum loops.
In contrast, the front-form (light-front) vacuum is defined as the lowest mass eigenstate of light-front Hamiltonian defined by quantizing at fixed $\tau = t -z/c$. The  vacuum is remarkably simple in light-front quantization because of the restriction $k^+ \ge 0.$   For example QED vacuum graphs such as $e^+ e^- \gamma $ do not arise.   The LF vacuum thus coincides with the  vacuum of the free LF Hamiltonian.  The front-form vacuum and  its eigenstates are causal and  Lorentz invariant; whereas the instant form vacuum depends on the observer's Lorentz frame.
The instant-form vacuum is a state defined at the same time $t$ at all spatial points in the universe.  In contrast, the front-from vacuum only senses phenomena which are causally connected; i.e., or within the observer's light-cone.
Causality in quantum field theory follows the fact that commutators vanish outside the light-cone. In fact in the LF analysis  the spatial support of QCD condensates
is restricted to the interior of hadrons, physics which arises due to the
interactions of confined quarks and gluons.
 In the Higgs theory, the usual Higgs vacuum expectation value is replaced with a $k^+=0$ zero mode;~\cite{Srivastava:2002mw} however, the resulting phenomenology is identical to the standard analysis.

When one makes a measurement in hadron physics, such as  deep inelastic lepton-proton scattering, one probes  hadron's constituents consistent with causality -- at a given light front time, not at instant time.  Similarly, when one makes observations in cosmology, information is obtained  within the causal horizon; i.e., consistent with the finite speed of light.
The cosmological constant measures the matrix element of the energy momentum tensor $T^{\mu \nu}$ in the background universe.  It corresponds to the measurement of the gravitational interactions of  a probe of finite mass;   it only senses the causally connected domain within the light-cone of the observer.  If the universe is empty, the appropriate vacuum state is thus the LF vacuum since it is causal.  One automatically obtains a vanishing cosmological constant from the LF vacuum.
Thus, as argued in Refs. ~\cite{Brodsky:2008be,Brodsky:2008xu}   the 45 orders of magnitude conflict of QCD with the observed value of the cosmological condensate is removed, and
a new perspective on the nature of quark and gluon condensates in
QCD  is thus obtained.~\cite{Brodsky:2008be,Brodsky:2008xu,Brodsky:2009zd}.

In fact, in the LF analysis one finds that the spatial support of QCD condensates
is restricted to the interior of hadrons, physics which arises due to the
interactions of color-confined quarks and gluons.  The condensate physics normally associated with the instant-form vacuum is replaced by the dynamics of higher non-valence Fock states as shown in the context of the infinite momentum/light-front method by Casher and Susskind.~\cite{Casher:1974xd}  and Burkardt~\cite{Burkardt:1997bd} In particular, chiral symmetry is broken in a limited domain of size $1/ m_\pi$,  in analogy to the limited physical extent of superconductor phases.
This novel description  of chiral symmetry breaking  in terms of ``in-hadron condensates''  has also been observed in Bethe-Salpeter studies.~\cite{Maris:1997hd,Maris:1997tm,Chang:2012rk,Chang:2011mu,Chang:2010jq}
The usual argument for a quark vacuum condensate is the Gell-Mann--Oakes--Renner
(GMOR)
formula:
$
m^2_\pi = -2 m_q {\langle0| \bar q q |0\rangle/ f^2_\pi}.
$
However, in the Bethe-Salpeter and light-front formalisms, where the pion is a $q \bar q$ bound-state, the GMOR relation is replaced by
$
m^2_\pi = - 2 m_q {\langle 0| \bar q \gamma_5  q |\pi \rangle/ f_\pi},
$
where $\rho_\pi \equiv - \langle0| \bar q \gamma_5  q |\pi\rangle$  represents a pion decay constant via an an elementary pseudoscalar current.
The result is independent of the renormalization scale. In the light-front formalism, this matrix element derives from the $|q \bar q \rangle$ Fock state of the pion with parallel spin-projections $S^z = \pm 1$ and $L^z= \mp 1$, which couples by quark spin-flip to the usual $|q \bar q\rangle$ $S^z=0, L^z=0$ Fock state via the running quark mass.   Thus again one finds ``in-hadron condensates"  replacing vacuum condensates: the $\langle 0|\bar q \bar q |0 \rangle$  vacuum condensate which appears in the Gell-Mann Oakes Renner
formula is replaced by the $\langle 0|\bar q \gamma_5|\pi \rangle$ pion decay constant.
This new perspective also explains the
results of studies~\cite{Ioffe:2002be,Davier:2007ym,Davier:2008sk} which find no significant signal for the vacuum gluon
condensate.

\section{Conclusions}

In this talk, I have  highlighted a number of areas where conventional wisdom in QCD and hadron physics has been  challenged.  These include standard assumptions such as

\begin{enumerate}

\item The heavy quark sea is conventionally assumed to arise only from gluon splitting and is thus confined to the low $x$ domain; in fact, QCD also predicts intrinsic contributions~\cite{Brodsky:1980pb} where the heavy quarks are multi-connected to the valence quarks  and appear at the same rapidity as the valence quarks; i.e., at large light-cone momentum fractions $x$.  This has important consequences for heavy quark phenomena at large $x_F$ and large transverse momentum as well as in weak decays of the $B$-meson~\cite{Brodsky:2001yt}.

\item  Initial-state and final-state Interactions are assumed to be  power-law suppressed. This is contradicted by factorization-breaking lensing phenomena such as the Sivers effect in polarized single-inclusive
deep inelastic  scattering~\cite{Brodsky:2002cx} as well as and the breakdown~\cite{Boer:2002ju} of the Lam-Tung relation in Drell-Yan reactions.

\item The structure function of a hadron is usually assumed to reflect just the physics of the wavefunction of the hadron, and thus it must be process independent; in fact, the observed structure functions are sensitive to process-dependent rescattering and lensing processes at leading twist.  One thus should distinguish dynamical versus static structure functions~\cite{Brodsky:2009dv}.

\item Antishadowing is a usually assumed to be a property of the nuclear wavefunction and is thus process-independent.   In fact as the NuTeV data shows~\cite{Schienbein:2008ay} each quark
can have its own antishadowing distribution~\cite{Brodsky:1989qz,Brodsky:2004qa}.

\item High-transverse momentum hadrons in inclusive reactions are usually assumed to arise only from jet fragmentation.  In fact, there is a significant probability for high $p_T$ hadrons  to arise from ``direct"  color-transparent subprocesses. This can explain anomalies in the fixed $x_T$ cross section and  the baryon anomaly, the large proton to pion ratio observed in heavy-ion collisions at RHIC~\cite{Arleo:2009ch}.

\item
Conventional wisdom states that the renormalization scale in QCD cannot be fixed and can only be guessed or chosen  to minimize sensitivity. In fact, it can be fixed at each order in perturbation theory using the principle of maximal conformality(PMC)~\cite{Brodsky:2011ig,Brodsky:2011ta}
The result is scheme-independent way and agrees with the conventional QED procedure in the Abelian limit.

\item It is standard wisdom that QCD condensates must be properties of the vacuum. In fact, one finds that vacuum condensates are replaced by hadronic matrix elements in the  Bethe-Salpeter and light-front analyses.   The conflict of traditional analyses with the cosmological constant highlights the need to distinguish different concepts of the vacuum: the acausal instant form vacuum versus the causal light-front definition~\cite{Brodsky:2010xf}.

\item Usually nuclei  are regarded as  composites of color-singlet nucleons; in fact, QCD predicts ``hidden color" configurations of the quarks which can dominate short distance nuclear reactions.~\cite{Ji:1985ky}

\item It is conventional to take the real part of  the virtual Compton scattering  amplitude to be  arbitrary subtraction term; in fact, local four-point photon-quark scattering leads to a real amplitude~\cite{Brodsky:2008qu}, a ``$J=0$ fixed pole" which is constant in energy and independent of the photons' virtuality at fixed $t$.

\item   Gluon degrees of freedom should be manifest at all scales - however, in AdS/QCD the effects of soft gluons are  sublimated in favor of the QCD confinement potential~\cite{Brodsky:2011pw}.

\item   Orbital angular momentum in the low lying hadrons is often assumed to be negligible. In fact, in AdS/QCD the nucleon eigensolutions for the light quarks have $L^z=\pm 1$ orbital components  comparable in strength to the $L^z=0$ component~\cite{Brodsky:2011zj}. This observation can help to explain the empirical fact that only a small fraction of the proton's spin is carried by quark or gluon spin.

\end{enumerate}

\begin{acknowledgements}
Presented at the International Symposium on Subnuclear Physics: Past, Present and Future, held at the Pontifical Academy of Sciences in the Vatican from 30 Oct. to 2 Nov. 2011,
I am grateful to Professor Antonio Zichichi, the organizer of the symposium, and Marcelo Sanchez Sorondo, the Chancellor  of the Pontifical Academy of Sciences at the Vatican, for their invitation to this unique and outstanding conference.   I also thank my collaborators for many helpful conversations.
This research was supported by the Department of Energy  contract DE--AC02--76SF00515.
SLAC-PUB-14865.

\end{acknowledgements}

\end{document}